\documentclass{article}

\usepackage{cite}
\usepackage[pdftex]{graphicx}
\usepackage{amsmath,amssymb,amsfonts}
\usepackage{algorithm}
\usepackage{algorithmic}
\usepackage{array}
\usepackage{subcaption}
\usepackage{overpic}

\usepackage{arxiv}

\usepackage{cite}
\usepackage[utf8]{inputenc} 
\usepackage[T1]{fontenc}    
\usepackage{hyperref}       
\usepackage{url}            
\usepackage{booktabs}       
\usepackage{amsfonts}       
\usepackage{nicefrac}       
\usepackage{microtype}      
\usepackage{lipsum}		
\usepackage{graphicx}
\usepackage{doi}

\usepackage{tikz}
\usepackage{textcomp}
\newcommand\copyrighttext{%
\footnotesize \textcopyright 2020 IEEE. Personal use of this material is permitted.
Permission from IEEE must be obtained for all other uses, in any current or future
media, including reprinting/republishing this material for advertising or promotional
purposes, creating new collective works, for resale or redistribution to servers or
lists, or reuse of any copyrighted component of this work in other works.

\doi{10.1109/RadarConf2043947.2020.9266412}}
\newcommand\copyrightnotice{%
\begin{tikzpicture}[remember picture,overlay]
\node[anchor=south,yshift=20pt] at (current page.south) {\fbox{\parbox{\dimexpr\textwidth-\fboxsep-\fboxrule\relax}{\copyrighttext}}};
\end{tikzpicture}%
}

\title{Near-Field MIMO-ISAR Millimeter-Wave Imaging}

\author{ \href{https://orcid.org/0000-0002-3388-4805}{\includegraphics[scale=0.06]{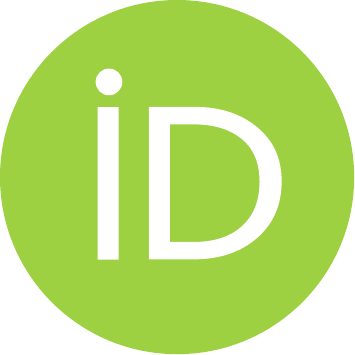}\hspace{1mm}Josiah W. Smith} \\
	Department of Electrical and Computer Engineering\\
	The University of Texas at Dallas\\
	Richardson, TX 75080 \\
	\texttt{josiah.smith@utdallas.edu} \\
	\And
	\href{https://orcid.org/0000-0001-8682-4577}{\includegraphics[scale=0.06]{orcid.pdf}\hspace{1mm}Muhammet Emin Yanik} \\
	Radar and Analytics\\
	Texas Instruments\\
	Dallas, TX 75243 \\
	\texttt{m-yanik@ti.com} \\
	\And
	\href{https://orcid.org/0000-0001-7229-1765}{\includegraphics[scale=0.06]{orcid.pdf}\hspace{1mm}Murat Torlak} \\
	Department of Electrical and Computer Engineering\\
	The University of Texas at Dallas\\
	Richardson, TX 75080 \\
	\texttt{torlak@utdallas.edu} \\
}

\date{}

\renewcommand{\headeright}{\textit{Proc. IEEE RadarConf} (Accepted)}
\renewcommand{\undertitle}{\textit{Proc. IEEE RadarConf} (Accepted)}
\renewcommand{\shorttitle}{Near-Field MIMO-ISAR}

\hypersetup{
pdftitle={Near-Field MIMO-ISAR Millimeter-Wave Imaging},
pdfsubject={eess.SP, cs.CV},
pdfauthor={Josiah W.~Smith, Muhammet Emin Yanik, Murat Torlak},
pdfkeywords={millimeter-wave (mmWave), multiple-input multiple-output (MIMO), inverse synthetic aperture radar (ISAR), three-dimensional (3-D) imaging},
}

\begin{document}
\maketitle
\copyrightnotice

\begin{abstract}
Multiple-input-multiple-output (MIMO) millimeter-wave (mmWave) sensors for synthetic aperture radar (SAR) and inverse SAR (ISAR) address the fundamental challenges of cost-effectiveness and scalability inherent to near-field imaging. In this paper, near-field MIMO-ISAR mmWave imaging systems are discussed and developed. The rotational ISAR (R-ISAR) regime investigated in this paper requires rotating the target at a constant radial distance from the transceiver and scanning the transceiver along a vertical track. Using a 77GHz mmWave radar, a high resolution three-dimensional (3-D) image can be reconstructed from this two-dimensional scanning taking into account the spherical near-field wavefront. While prior work in literature consists of single-input-single-output circular synthetic aperture radar (SISO-CSAR) algorithms or computationally sluggish MIMO-CSAR image reconstruction algorithms, this paper proposes a novel algorithm for efficient MIMO 3-D holographic imaging and details the design of a MIMO R-ISAR imaging system. The proposed algorithm applies a multistatic-to-monostatic phase compensation to the R-ISAR regime allowing for use of highly efficient monostatic algorithms. We demonstrate the algorithm's performance in real-world imaging scenarios on a prototyped MIMO R-ISAR platform. Our fully integrated system, consisting of a mechanical scanner and efficient imaging algorithm, is capable of pairing the scanning efficiency of the MIMO regime with the computational efficiency of single pixel image reconstruction algorithms.
\end{abstract}

\keywords{millimeter-wave (mmWave) \and multiple-input multiple-output (MIMO) \and inverse synthetic aperture radar (ISAR) \and three-dimensional (3-D) imaging}

\section{Introduction}
\label{sec:introduction}
Over the past several decades, developments in system-on-chip complementary metal oxide semiconductor (CMOS) radio frequency integrated circuits (RFIC) have resulted in the emergence of frequency modulated continuous wave (FMCW) millimeter wave (mmWave) radars as a cost-effective solution for imaging applications. The 3-D holographic imaging regime has been investigated in the rectilinear (planar) mode \cite{yanik2018millimeter,qiao2015compressive} and in cylindrical mode \cite{sheen2010near}. Additionally, progress has been made towards efficient algorithms for single-input-single-output (SISO) monostatic array synthetic aperture radar (SAR) \cite{sheen2001three} and multi-input-multi-output (MIMO) multistatic array SAR \cite{zhuge2012three}. Specifically, Gao's work at China's National University of Defense and Technology (NUDT) has demonstrated algorithms for 2-D circular SAR (CSAR) imaging \cite{gao2016efficient} and 3-D MIMO-CSAR imaging \cite{gao2018cylindricalMIMO}. While SISO-CSAR algorithms proposed by Sheen \cite{sheen1999real}, Laviada \cite{laviada2017multiview}, Gao, and others are efficient in generating high resolution 3-D holographic images, they ignore the multistatic effects from a MIMO array, resulting in aliasing and phase mismatch from the ideal SISO case. While MIMO-CSAR algorithms have been developed in attempt to solve such issues, these algorithms are computationally expensive and inefficient in comparison to their SISO counterparts. In this paper, we propose a resolution to this dilemma by leveraging the benefits of MIMO-CSAR, fewer antenna elements and cost efficiency, with the streamlined computational efficiency of the SISO-CSAR algorithms to produce a highly efficient high-resolution 3-D imaging algorithm. Under this MIMO rotation ISAR (R-ISAR) regime, a robust imaging system is prototyped to verify the proposed algorithm and demonstrate its performance.

The rest of this paper is formatted as follows. Section \ref{sec:signal_model} discusses the return signal from the  proposed MIMO R-ISAR scenario and the multistatic-to-monostatic conversion. Section \ref{sec:derivation_3D_siso_algorithm} contains the derivation for the 3-D image reconstruction algorithm in the SISO R-ISAR regime and crucial multistatic-to-monostatic phase correction. Section \ref{sec:key_issues} overviews issues including sampling criteria and spatial resolution. Section \ref{sec:simulations} verifies the proposed algorithm in simulation. The imaging prototype is described in Section \ref{sec:experimental_setup}. Real 3-D imaging results are reported in Section \ref{sec:imaging_results}, including a comparison of R-ISAR to SAR, followed finally by conclusions.

\begin{figure}[h]
	\centering
	\includegraphics[width=3.4in]{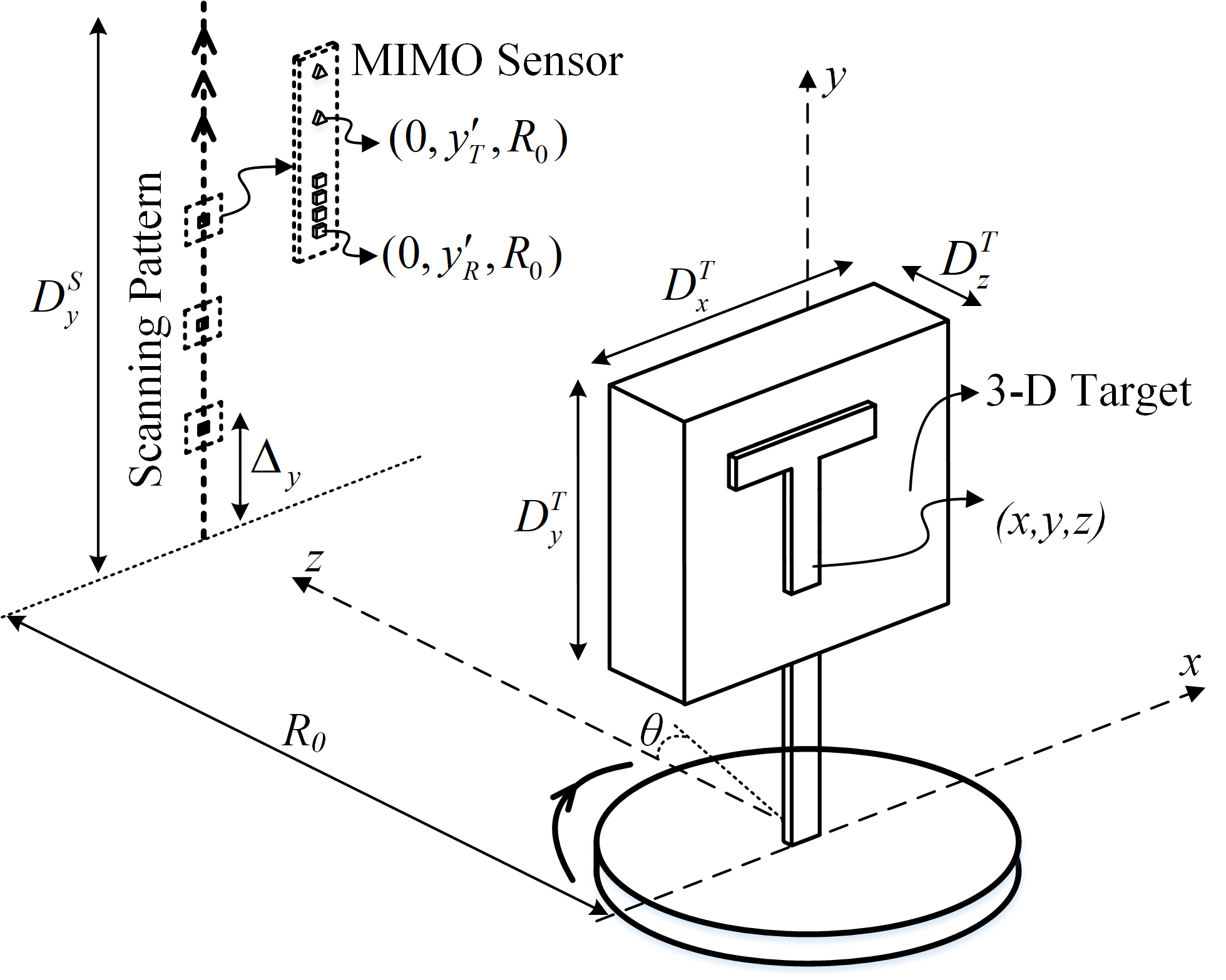}
	\caption{The geometry of the MIMO R-ISAR imaging configuration, where a cylindrical aperture is synthesized by mechanically moving a linear MIMO array vertically and rotating the target.}
	\label{fig:MIMO_R_ISAR_System_Configuration}
\end{figure}

\section{MIMO R-ISAR Signal Model}
\label{sec:signal_model}

\subsection{MIMO Rotational ISAR (R-ISAR) Echo Signal}
\label{Sec_mimo_r-isar_signal}

The MIMO R-ISAR scenario, as shown in Fig. 1, consists of a rotational scanner whose center is the origin and a MIMO array scanned along the y-axis (vertically), located at a constant distance of $R_0$ from the center of the rotator. The distance of the transmitter, located at the vertical position $y_T'$, and receiver, located at the vertical position $y_R'$, from each point in the target domain $(x,y,z)$ depends on the rotation angle $\theta$ and the distance $R_0$.

\begin{align}
\begin{split}
R_T = \sqrt{(x - R_0 \cos\theta)^2 + (z - R_0 \sin\theta)^2 + (y - y_T')^2}, \\
R_R = \sqrt{(x - R_0 \cos\theta)^2 + (z - R_0 \sin\theta)^2 + (y - y_R')^2}.
\end{split}
\end{align}

The MIMO echo signal from the R-ISAR scenario can be modeled as:
\begin{equation}
\label{Eq_mimo_echo_signal}
\text{s}(\theta,k,y_T',y_R') = \iiint \frac{\text{p}(x,y,z)}{R_T R_R} e^{jk(R_T + R_R)} dx dy dz,
\end{equation}
where $\theta$ is the angle of rotation, and $k$ is the wavenumber. For our purposes $\theta$ is allowed to be wide-angle up to $360^{\circ}$.

\subsection{Multistatic-to-Monostatic Conversion}
\label{Sec_multistatic_to_monostatic_conversion}

The received echo data from the multistatic MIMO array undergo a simple transformation to approximate its counterpart echo signal from the virtual-SISO array elements. To convert this four-dimensional (4-D) MIMO signal to a 3-D virtual-SISO signal, the following phase compensation is performed \cite{yanik2019cascaded,yanik2019sparse},

\begin{equation}
\label{Eq_phase_compensation}
\hat{\text{s}}(\theta,k,y') = \text{s}(\theta,k,y_T',y_R') e^{-jk\frac{Sy^2}{4R_0}},
\end{equation}
where $y'$ is the vertical scanning dimension containing all $y_T'$ and $y_R'$ positions, and $d_y$ is the distance between pairs of transmitting and receiving elements. This multistatic-to-monostatic conversion only holds for small values of $d_y$.

This compensation is a crucial step in the algorithm. By compensating the phase of the multistatic MIMO signal to obtain an approximate of the echo signal from virtual elements located at the midpoint of each MIMO transmitter/receiver pair, this virtual-SISO data can be fed into the efficient 3-D imaging algorithm derived in section \ref{sec:derivation_3D_siso_algorithm}.

\section{Derivation of 3-D Image SISO Reconstruction Algorithm}
\label{sec:derivation_3D_siso_algorithm}
	
The derivations given in this section are similar to the CSAR algorithm derived by Sheen in \cite{sheen1999real}, but contain several key differences vital to performing successful 3-D holographic image reconstruction for a circular scanning scenario.

Using the R-ISAR scenario shown in Fig. 1, the return signal from a monostatic SISO transceiver, neglecting amplitude terms, can be modeled as

\begin{equation}
\hat{\text{s}}(\theta,k,y') = \iiint \text{p}(x,y,z) e^{j2kR} dx dy dz,
\label{Eq_SISO_echo_signal}
\end{equation}
where

\begin{equation}
R = \sqrt{(x - R_0\cos\theta)^2 +  (z - R_0\sin\theta)^2 + (y - y')^2},
\end{equation}
and $\text{p}(x,y,z)$ is the complex reflectivity function of the target scene. Using the method of stationary phase (MSP), the exponential term in (\ref{Eq_SISO_echo_signal}) can be decomposed by

\begin{align}
\begin{split}
&e^{j2k\sqrt{(R_0\cos\theta - x)^2 + (R_0\sin\theta - z)^2 + (y - y')^2}} = \\ 
&\iint e^{jk_r\cos\phi(R_0\cos\theta - x) + jk_r\sin\phi(R_0\sin\theta - z) + jk_{y'}(y-y'))} d\phi dk_{y'}.
\end{split}
\label{Eq_MSP_step}
\end{align}

Note that an identical result can be found by decomposing the free-space Green's function of a point source in the spatial spectral domain \cite{detlefsen2005effective}. The angle of each plane wave component in the $x\text{-}z$ plane is $\phi$, and $k_{y'}$ is the y-component of the wavenumber. Using the dispersion relation

\begin{equation}
4k^2 = k_x^2 + k_y^2 + k_z^2,
\end{equation}
we define $k_r$ as the wavenumber component in the $x\text{-}z$ plane as

\begin{equation}
\label{Eq_stolt_1}
k_r = \sqrt{k_x^2+k_z^2} = \sqrt{4k^2 - k_y^2}.
\end{equation}

Combining the above relations yields
\begin{align}
\begin{split}
\hat{\text{s}}(&\theta,k,y') = \\
& \iint \left[ \iiint \text{p}(x,y,z)e^{-j(k_r\cos\phi)x-j(k_r\sin\phi)z-jk_{y'}y}dxdydz \right] \\
& \times e^{jk_rR_0\cos(\theta-\phi) + jk_{y'}y'} d\phi dk_{y'},
\end{split}
\label{Eq_mimo_r_isar_signal_expanded}
\end{align}

The term inside the $[\bullet]$ brackets is the 3-D Fourier transform of the reflectivity function. Using the following Fourier transform pair in polar coordinates:

\begin{equation}
\text{p}(x,y,z) \iff \text{P}(k_r\cos\phi,k_y,k_r\sin\phi),
\end{equation}
(\ref{Eq_mimo_r_isar_signal_expanded}) yields
\begin{align}
\begin{split}
\hat{\text{s}}(\theta,k,y') &= \iint e^{jk_rR_0\cos(\theta-\phi)} \\
&\times \text{P}(k_r\cos\phi,k_y,k_r\sin\phi) e^{jk_{y'}y'}d\phi dk_{y'}.
\end{split}
\label{Eq_mimo_r_isar_signal_expanded_2}
\end{align}

Taking the Fourier transform with respect to $y'$ on both sides and dropping the distinction between $y'$ and $y$ due to coincidence of the domains:

\begin{equation}
\hat{\text{S}}(\theta,k,k_y) = \int_{-\frac{\pi}{2}}^{\frac{\pi}{2}} e^{jk_rR_0\cos(\theta-\phi)} \text{P}(k_r\cos\phi,k_y,k_r\sin\phi)d\phi. 
\end{equation}

Defining:

\begin{align}
\label{Eq_mimo_r_isar_signal_expanded_3}
\hat{\text{P}}(\phi,k_r,k_y) &\triangleq \text{P}(k_r\cos\phi,k_y,k_r\sin\phi) \\
\text{g}(\theta,k_r) &\triangleq e^{jk_rR_0\cos\theta}.
\end{align}

Now:

\begin{equation}
\hat{\text{S}}(\theta,k,k_y) = \int_{-\frac{\pi}{2}}^{\frac{\pi}{2}} \text{g}(\theta-\phi,k_r) \hat{\text{P}}(\phi,k_r,k_y)d\phi, 
\end{equation}
which represents a convolution in the $\theta$ domain:

\begin{equation}
\hat{\text{S}}(\theta,k,k_y) = \text{g}(\theta,k_r) \circledast_\theta \hat{\text{P}}(\theta,k_r,k_y),
\end{equation}
where $\circledast_\theta$ is the convolution operator along the $\theta$ domain.

Taking the Fourier transform across the $\theta$ domain on both sides yields:

\begin{equation}
\hat{\text{S}}(\Theta,k,k_y) = \text{G}(\Theta,k_r)\tilde{\text{P}}(\Theta,k_r,k_y),
\end{equation}
where

\begin{align}
\text{G}(\Theta,k_r) &= \text{FT}_{1D}^{(\theta)}[\text{g}(\theta,k_r)],\\
\tilde{\text{P}}(\Theta,k_r,k_y) &= \text{FT}_{1D}^{(\theta)}[\hat{\text{P}}(\theta,k_r,k_y)]
\end{align}

Solving for $\tilde{P}$ by taking the inverse filter $G^*(\Theta,k_r)$ and then taking an inverse Fourier transform across the $\Theta$ domain for both sides to obtain $\hat{P}(\theta,k_r,k_y)$:

\begin{align}
\tilde{\text{P}}(\Theta,k_r,k_y) &= \hat{\text{S}}(\Theta,k,k_y)\text{G}^*(\Theta,k_r), \\
\hat{\text{P}}(\theta,k_r,k_y) &= \text{IFT}_{1D}^{(\Theta)}\left[ \hat{\text{S}}(\Theta,k,k_y)\text{G}^*(\Theta,k_r) \right]
\end{align}

By (\ref{Eq_mimo_r_isar_signal_expanded_3}):

\begin{equation}
\text{P}(k_r \cos\theta,k_r \sin\theta,k_y) = \text{IFT}_{1D}^{(\Theta)}\left[ \hat{\text{S}}(\Theta,k,k_y)\text{G}^*(\Theta,k_r) \right],
\end{equation}
where $k_x = k_r \cos\theta$ and $k_z = k_r \sin\theta$. The definition of these horizontal wavenumber components is crucial in the derivation of the reconstruction algorithm and have been improperly defined in past MSP-based algorithms \cite{sheen1999real}. $\hat{P}(\theta,k_r,k_y)$ will be a uniformly sampled function of $\theta$ and $k_r$ and will need to be interpolated onto a uniform $(k_x,k_z,k_y)$ grid via Stolt interpolation using the equation (\ref{Eq_stolt_1}) and the following relations:

\begin{align}
\label{Eq_stolt_2}
\theta &= \tan^{-1}\left(\frac{k_z}{k_x}\right), \\
\label{Eq_stolt_3}
k &= \frac{1}{2}\sqrt{k_x^2 + k_y^2 + k_z^2}.
\end{align}

The Stolt interpolation process will be denoted by the $\mathcal{S}[\bullet]$ operator, such that:
\begin{equation}
\text{P}(k_x,k_y,k_z) = \mathcal{S}[\text{P}(k_r \cos\theta,k_r \sin\theta,k_y)].
\end{equation}

Finally, the algorithm can be summarized by (\ref{Eq_risarAlgoSummary_1}) and (\ref{Eq_risarAlgoSummary_2}).
\begin{align}
\label{Eq_risarAlgoSummary_1}
\text{p}(x,y,z) &= \text{IFT}_{3D}^{(k_x,k_y,k_z)}\left[ \text{P}(k_x,k_y,k_z) \right], \\
\label{Eq_risarAlgoSummary_2}
\text{P}(k_x,k_y,k_z) &= \mathcal{S}\left[\text{IFT}_{1D}^{(\Theta)}\left[ \text{FT}_{2D}^{(\theta,y)}\left[\hat{\text{s}}(\theta,k,y)\right]\text{G}^*(\Theta,k_r)\right]\right].
\end{align}

From the above results, the complete 3-D image reconstruction algorithm is summarized below. To our knowledge, the key pairing of multistatic-to-monostatic conversion with a single pixel reconstruction algorithm has not been shown in prior literature. This novelty allows for an efficient imaging system by leveraging a MIMO array topology without increasing the computational complexity. In that way, the benefits of the MIMO virtual array can be achieved without the need for inefficient MIMO reconstruction algorithms. Rather, with the proposed algorithm, an efficient system can be easily implemented to quickly scan a target and immediately reproduce a high-resolution 3-D holographic image, as discussed in Section \ref{sec:imaging_results}.

\textbf{Efficient MIMO R-ISAR 3-D Holographic Imaging Algorithm}
\begin{enumerate}
	\item Gather the raw 4-D MIMO echo data as $\text{s}(\theta,k,y_T,y_R)$.
	\item Perform the phase compensation described in (\ref{Eq_phase_compensation}) to acquire $\hat{\text{s}}(\theta,k,y)$, the 3-D monostatic equivalent.
	\item Perform a 2-D FFT across the $\theta$ and $y$ dimensions of the phase corrected data to obtain $\hat{S}(\Theta,k,k_y)$.
	\item Generate the azimuth filter $\text{g}(\theta,k_r) \triangleq e^{jk_rR_0\cos\theta}$ and implement an FFT across the $\theta$ dimension to compute the spectral azimuth filter $\text{G}(\Theta,k_r)$
	\item Multiply $\hat{\text{S}}(\Theta,k,k_y)$ by the inverse filter $\text{G}^*(\Theta,k,k_y)$  and perform an IFFT across the $\Theta$ domain to obtain $\text{P}(k_r\cos\theta,k_r\sin\theta,k_y)$.
	\item Apply Stolt interpolation using the relations in (\ref{Eq_stolt_1}), (\ref{Eq_stolt_2}), and (\ref{Eq_stolt_3}) to transform the polar spatial spectral $\text{P}(k_r\cos\theta,k_r\sin\theta,k_y)$ to the Cartesian $\text{P}(k_x,k_y,k_z)$.
	\item Finally, compute a 3-D IFFT across $k_x$, $k_y$, and $k_z$ to recover the complex reflectivity function $\text{p}(x,y,z)$.
\end{enumerate}

\section{Discussion of Key Imaging Issues}
\label{sec:key_issues}

\subsection{Sampling Criteria}
Akin to all sampling applications, spatial sampling in the R-ISAR regime must satisfy the spatial Nyquist theorem. Accordingly, the following sampling criteria must be satisfied for alias-free 3-D holographic image reconstruction as discussed in \cite{gao2018cylindricalMIMO,sheen1999real,zhuge2010sparse}.

\begin{align}
	\Delta_k &< \frac{\pi}{2R_T}, \\
	\Delta_y &< \frac{\lambda \sqrt{(D_y^S + D_y^T)^2 /4 + R_0^2}}{2(D_y^S + D_y^T)}, \\
	\Delta_\theta &< \frac{\pi \sqrt{R_0^2 + R_T^2}}{2k_{max}R_0R_T}.
\end{align}
$R_T$ is the maximum radius of the target scene, $k_{max}$ is the maximum wavenumber, and $D_y^T$ and $D_y^S$ are the target and scan height, respectively.

\subsection{Spatial Resolution}
Another significant point of discussion for SAR imaging systems is spatial resolution. Vertical resolution is independent of the horizontal rotation and can be calculated using the effective aperture approach as shown in \cite{zhuge2010sparse}. 
\begin{equation}
	\delta_y \approx \frac{\lambda_c R_0}{2D_y^S}
\end{equation}
$\lambda_c$ is the wavelength of the center frequency. 
To derive the radial resolution, the problem is restricted to a 2-D horizontal plane, thereby removing the vertical element of the scan. Along this horizontal plane, the point spread function (PSF) can be computed analytically as \cite{gao2016terahertz}:

\begin{equation}
	\label{Eq_2D_PSF}
	\text{PSF}(r,\theta) = k_{max} \frac{\text{J}_1(2k_{max}r)}{\pi r} - k_{min} \frac{\text{J}_1(2k_{min}r)}{\pi r}
\end{equation}

From (\ref{Eq_2D_PSF}), the horizontal resolution can be deduced, where $\text{J}_1(\bullet)$ represents the first-order Bessel function and $k_{min}$ is the minimum wavenumber. 

\begin{equation}
	\delta_R = \frac{2.4}{k_{max} + k_{min}}
\end{equation}

For both the vertical and horizontal resolutions, an ideal point reflector is employed for simplicity sake. However, for real-world applications involving real target scenes, this type of ideal spatial resolvability is rarely achieved \cite{gao2018cylindricalMIMO}. Accordingly, these expressions serve as a lower limit on the empirical spatial resolution.

\section{R-ISAR Simulations}
\label{sec:simulations}

To simulate the echo signal, targets are modeled as point reflectors using  (\ref{Eq_mimo_echo_signal}). All the simulations are done using the MIMO parameters shown in Table \ref{table_radar_parameters} where $\theta_{max}$ is the maximum rotation angle, $N_y$ is the number of vertical captures, and $\Delta_y$ is the vertical spacing between MIMO captures.

\subsection{Point Spread Function (PSF)}
	
\begin{figure} [h]
	\begin{subfigure}{.5\linewidth}
		\centering
		\includegraphics[width=1\linewidth]{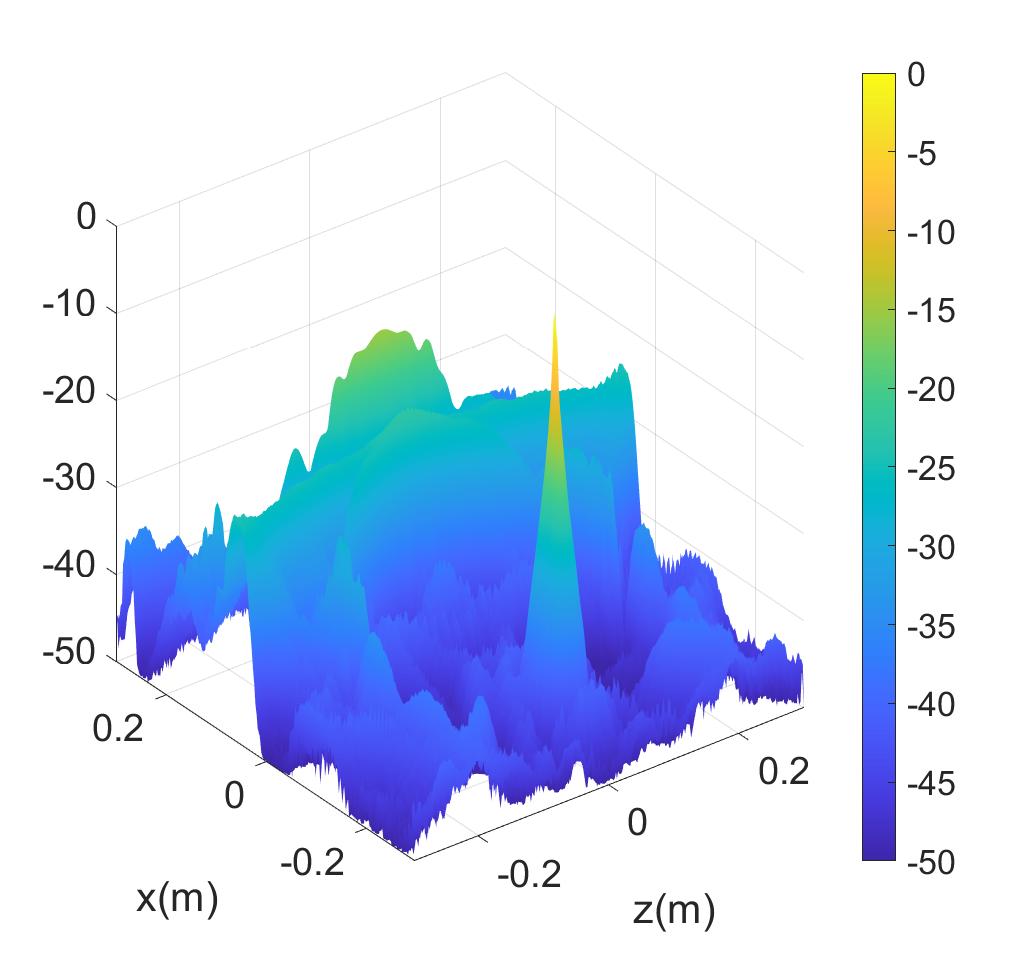}
		\caption{}
		\label{fig:psf_xz_R0_250_DS_478}
	\end{subfigure}%
	\begin{subfigure}{.5\linewidth}
		\label{fig:psf_xy_R0_250_DS_478}
		\centering
		\includegraphics[width=1\linewidth]{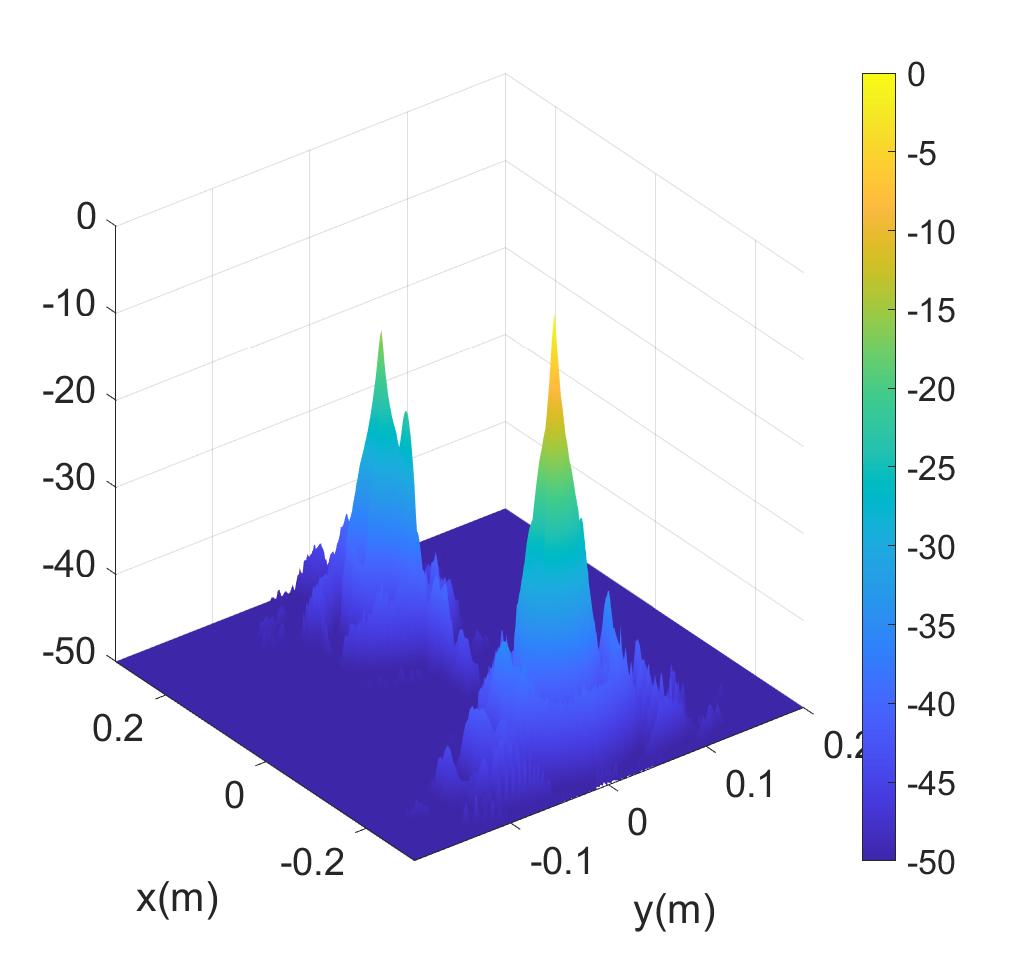}
		\caption{}
	\end{subfigure}
	\caption{Point Spread Function (dB): $R_0$ = 0.25 m and $D^S_y$ = 484.8 mm (a) 2-D x-z PSF, (b) 2-D x-y PSF}
	\label{fig:psf}
\end{figure}

Using (\ref{Eq_mimo_echo_signal}), the echo signal is simulated in MATLAB with a single point reflector located at $(-0.15,0)$ in the $x\text{-}z$ plane and with the parameters in Table \ref{table_radar_parameters}. Then, the image is reconstructed using the proposed algorithm described in Section \ref{sec:derivation_3D_siso_algorithm}. 2-D slices of the 3-D PSF are examined in Fig. \ref{fig:psf}.

\subsection{3-D Points}
\begin{figure} [h]
	\begin{subfigure}{.5\linewidth}
		\centering
		\includegraphics[width=1\linewidth]{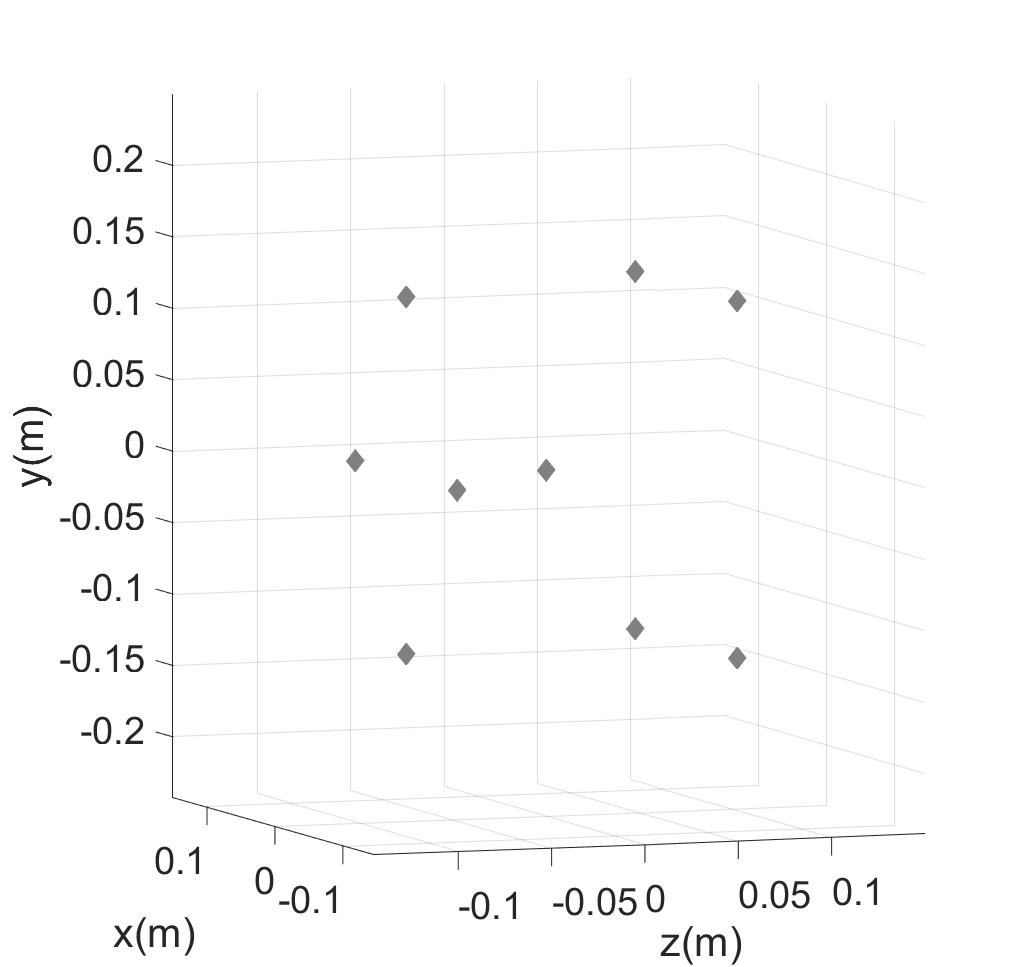}
		\caption{}
		\label{fig:grid3D_in}
	\end{subfigure}%
	\begin{subfigure}{.5\linewidth}
		\label{fig:grid3D_out}
		\centering
		\includegraphics[width=1\linewidth]{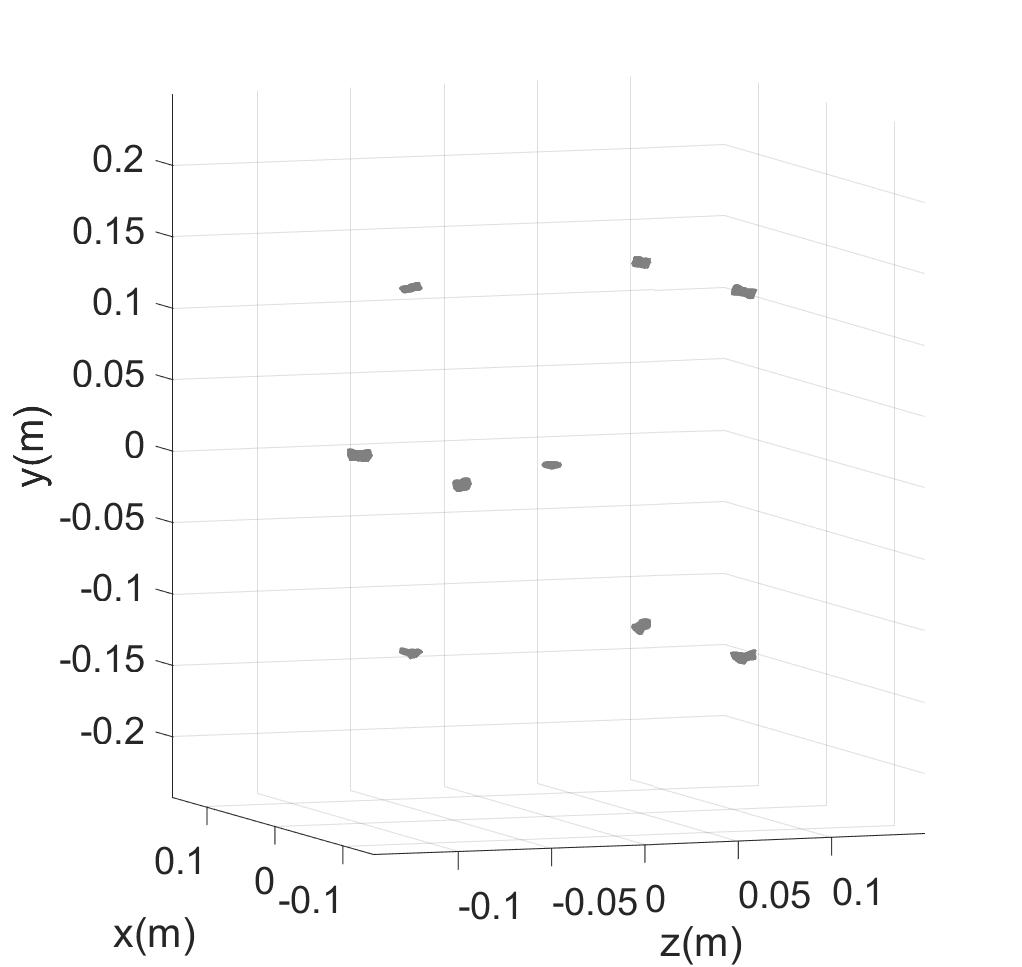}
		\caption{}
	\end{subfigure}
	\caption{(a) 3-D input grid of points reflectivity function, and (b) 3-D reconstructed image.}
	\label{fig:simulation_grid}
\end{figure}

Additionally, to verify the algorithm in simulation, a set of points in 3-D are generated and their echo signal is simulated. The algorithm again effectively reconstructs the images producing a nearly perfect duplicate of the input reflectivity function as shown in Fig. \ref{fig:simulation_grid}.

With successful verification of the algorithm in simulation, a custom prototype R-ISAR scanner is built to experimentally capture data and test the algorithm's image quality on real echo data.

\section{Experimental Setup}
\label{sec:experimental_setup}

\begin{figure} [h]
	\begin{subfigure}{.5\linewidth}
		\centering
		\includegraphics[width=1\linewidth]{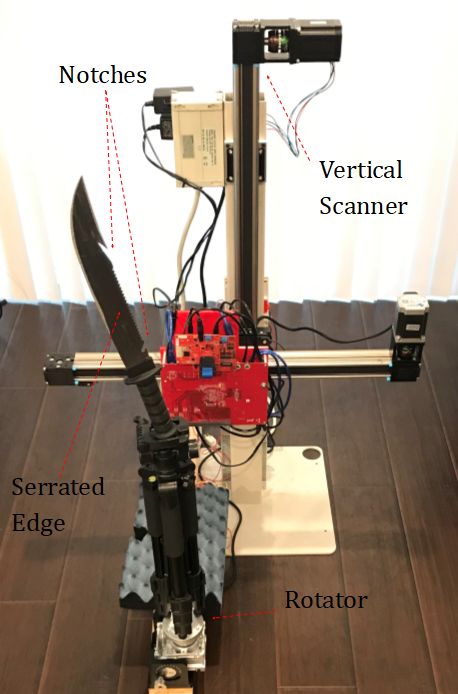}
		\caption{}
		\label{fig:scanner}
	\end{subfigure}%
	\begin{subfigure}{.5\linewidth}
		\centering
		\includegraphics[width=1\linewidth]{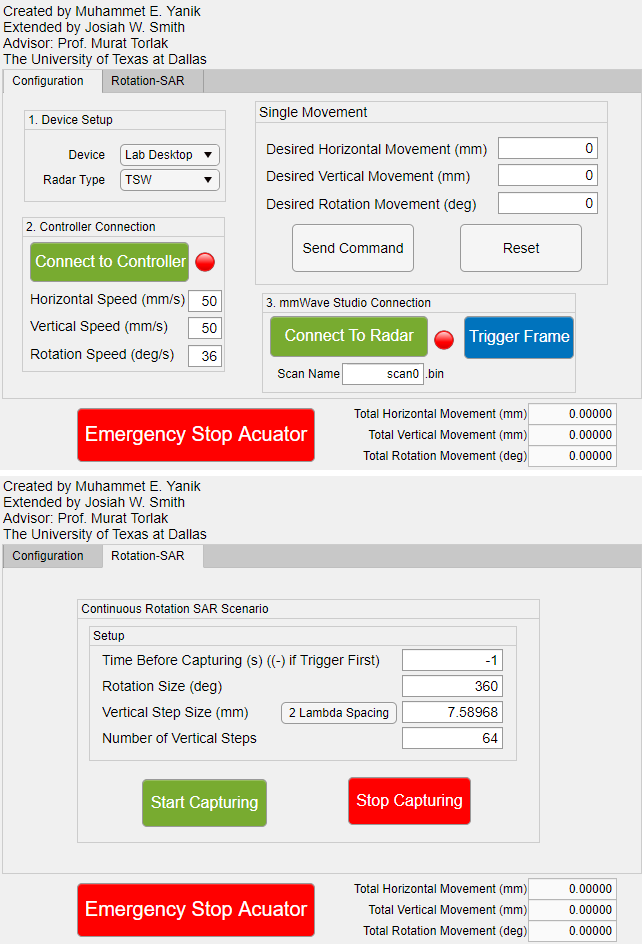}
		\caption{}
		\label{fig:matlab_gui}
	\end{subfigure}
	\caption{(a) Custom-built rotation-ISAR scanner, and (b) MATLAB GUI.}
\end{figure}

A cylindrical aperture is synthesized by mechanically moving a linear MIMO array continuously along a vertical track pattern, and rotating the target as shown in Fig.~\ref{fig:MIMO_R_ISAR_System_Configuration}. The system consists of Texas Instruments (TI) IWR1443-Boost, mmWave-Devpack, and TSW1400 mounted on a 2-D vertical and horizontal scanner as shown in Fig. \ref{fig:scanner}. For this application, only the vertical motion is used. The scanned object is mounted on the rotator. All mechanical motions are controlled by stepper drivers and embedded microcontrollers. The entire setup is controlled by a custom MATLAB graphical user interface (GUI), shown in Fig. \ref{fig:matlab_gui}. 

\begin{table} [h]
	\caption{\label{table_radar_parameters}R-ISAR Radar Parameters}
	\centering
	\begin{tabular}{c| c c c c c c}
		\hline
		& $R_0$ & $\Delta_\theta$ & $\theta_{max}$ & $N_y$ & $D^S_y$ & $\Delta_y$  \\ [0.5ex] 
		\hline\hline
		MIMO & 0.25 m & $0.036^{\circ}$ & $360^{\circ}$ & 64 & 484.8 mm  & $2\lambda$  \\ 
		\hline
		SISO & 0.25 m & $0.036^{\circ}$ & $360^{\circ}$ & 512 & 484.8 mm & $\lambda/4$ \\ 
		\hline
	\end{tabular}
\end{table}

\section{Imaging Results}
\label{sec:imaging_results}
The 2-D vertical and rotational scan is performed by the prototype scanner. The large knife shown in Fig. \ref{fig:scanner} is mounted to the rotator at an angle and scanned. Note the knife's notches and serrated edge. 

\subsection{SISO Imaging Results}

In the first experiment, a single transceiver pair is used to simulate a full-duplex SISO transceiver using the full 4 GHz bandwidth. Since the MIMO virtual array consists of 8 equally spaced virtual elements spanning $2\lambda$, the SISO scan will have a vertical spacing of $\lambda/4$ and will require 512 vertical captures to replicate the MIMO scan, drastically increasing the scanning time. All other parameters will remain the same, as shown in Table \ref{table_radar_parameters}.

Neglecting the multistatic-to-monostatic conversion, the proposed algorithm is implemented on the SISO echo data to produce a holographic image, as shown in Fig. \ref{fig:knife_SISO_mip}. While the knife's intricacies are clearly visible in the high-resolution image, the entire scan took nearly two and a half hours to complete. Next, we exploit the MIMO virtual array to drastically reduce the scanning time. The key novelty of the proposed algorithm is the increase in scanning efficiency by leveraging a MIMO topology without increasing the computational complexity.

\subsection{MIMO Imaging Results}

\begin{figure} [h]
	\begin{subfigure}{.45\linewidth}
		\centering
		\includegraphics[width=1\linewidth]{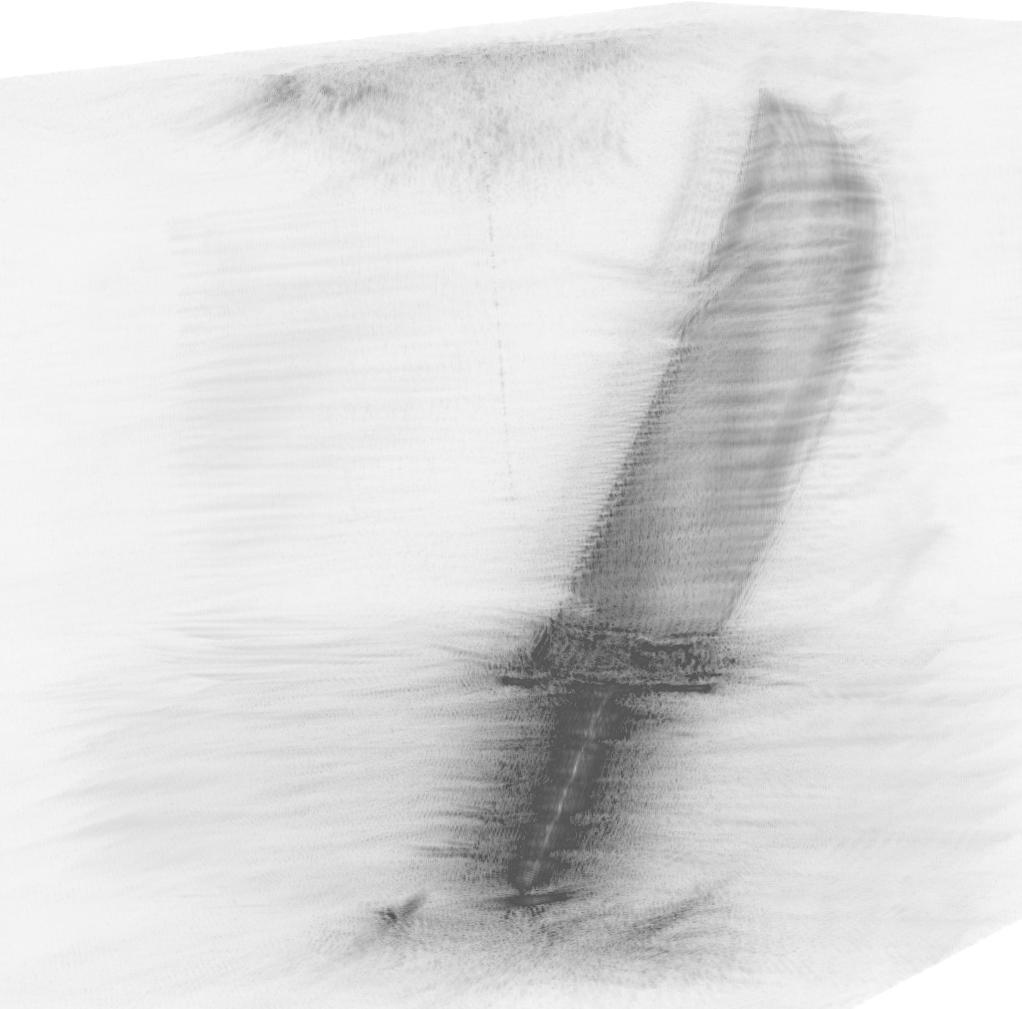}
		\caption{}
		\label{fig:knife_SISO_mip}
	\end{subfigure}%
	\begin{subfigure}{.45\linewidth}
		\centering
		\includegraphics[width=1\linewidth]{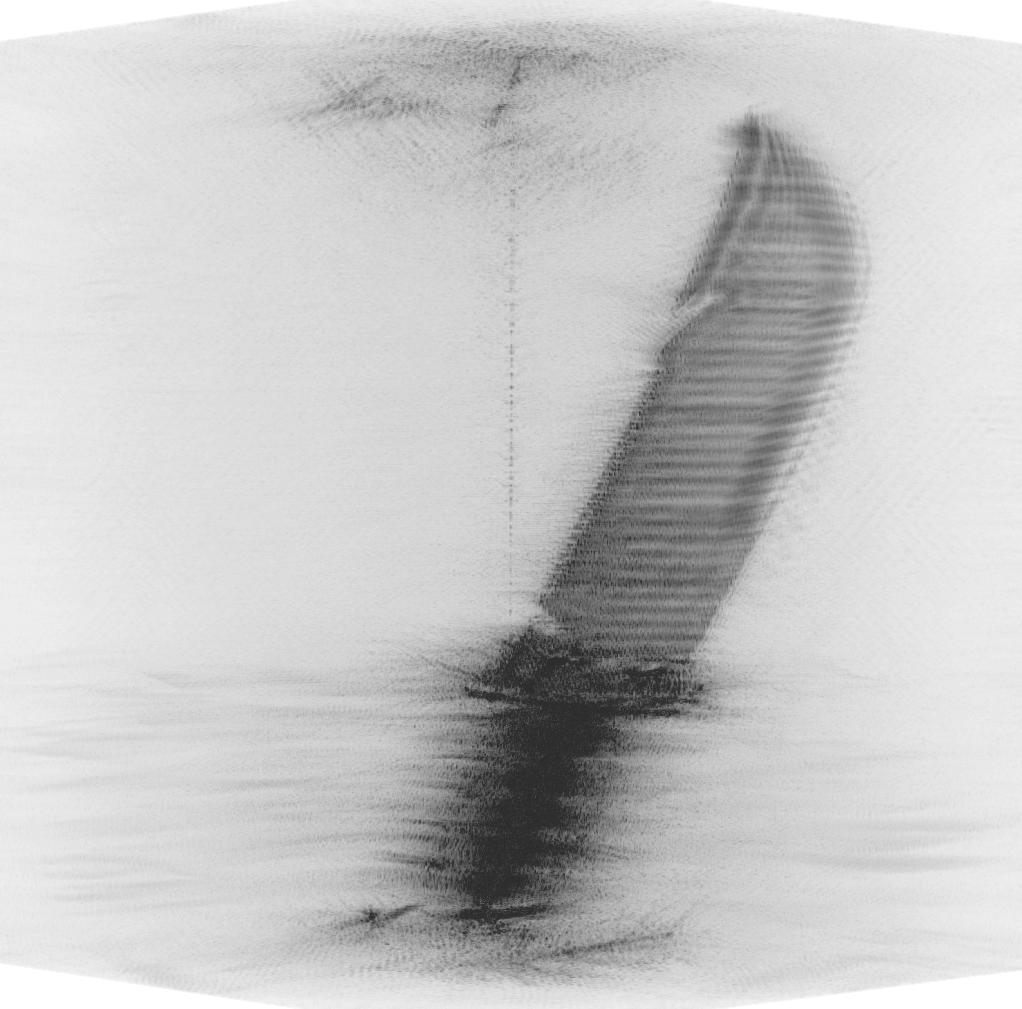}
		\caption{}
		\label{fig:knife_MIMO_mip}
	\end{subfigure}
	\begin{subfigure}{.45\linewidth}
		\centering
		\includegraphics[width=1\linewidth]{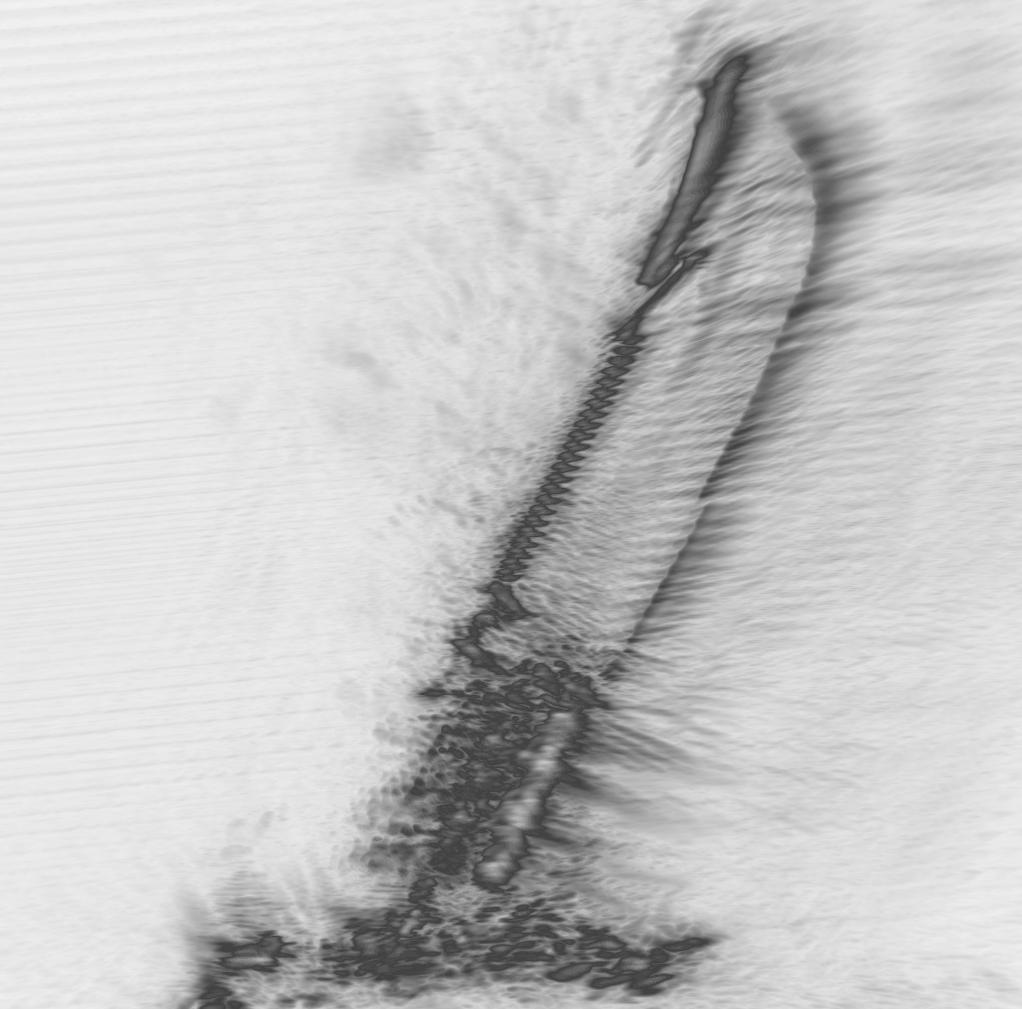}
		\caption{}
		\label{fig:knife_MIMO_SAR_Parallel_mip}
	\end{subfigure}%
	\begin{subfigure}{.45\linewidth}
		\centering
		\includegraphics[width=1\linewidth]{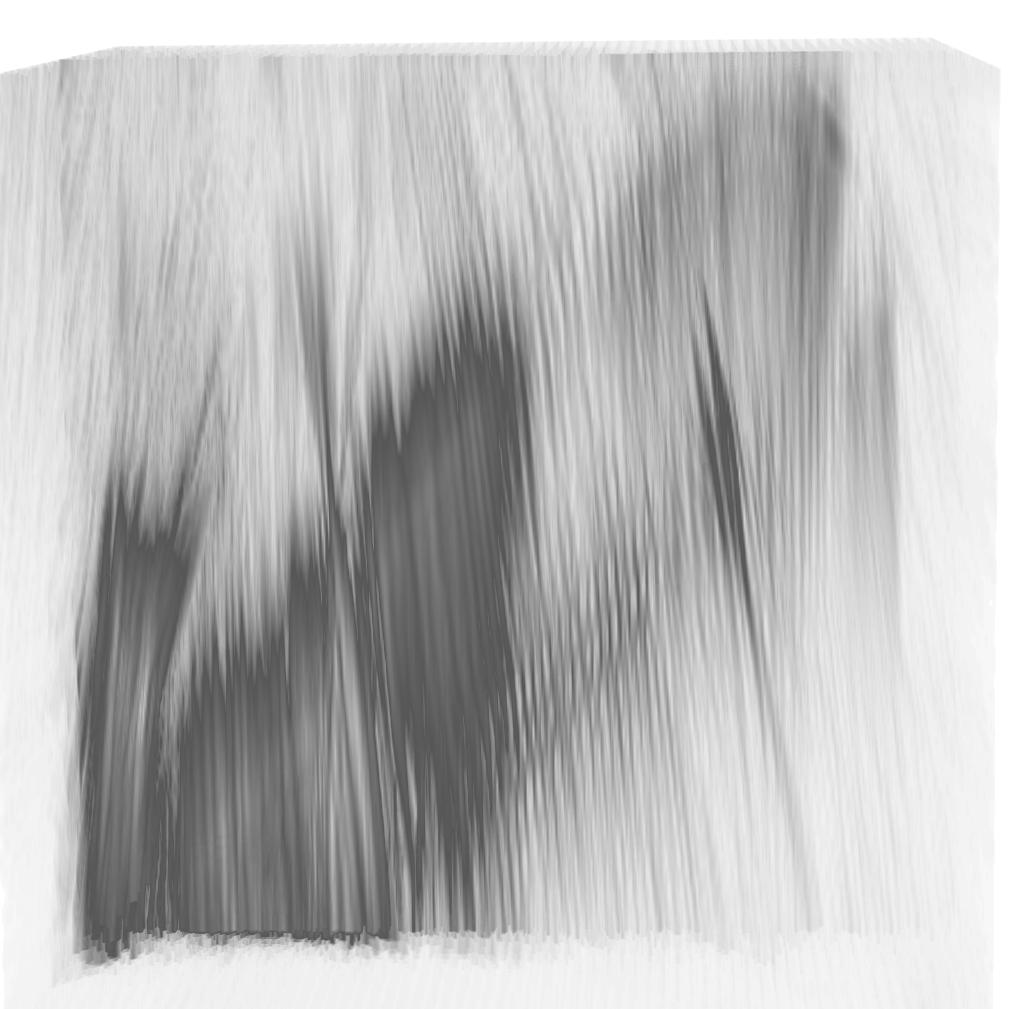}
		\caption{}
		\label{fig:knife_MIMO_SAR_Perp_mip}
	\end{subfigure}
	\caption{Comparison Reconstructed Images: (a) SISO R-ISAR 3-D MIP, (b) MIMO R-ISAR 3-D MIP, (c) MIMO R-SAR with knife parallel to scanning plane, (d) MIMO R-SAR with knife perpendicular to scanning plane}
	\label{fig:knife_MIMO}
\end{figure}

Now, the knife is scanned again, this time using 2 transmitters and 4 receiver antennas on the TI IWR1443-Boost and again 4 GHz bandwidth. After calibration to remove instrument delay, the echo data is processed by the proposed R-ISAR MIMO 3-D holographic imaging algorithm and an image is produced. Again, a 3-D rendering is included below, in Fig. \ref{fig:knife_MIMO_mip}. The MIMO scan only took less than twenty minutes to complete, a significant reduction in comparison to the SISO scan without degrading the image quality.

\subsection{Rectilinear MIMO-SAR Comparison}
\begin{figure} [h]
	\begin{subfigure}{.5\linewidth}
		\centering
		\includegraphics[width=1\linewidth]{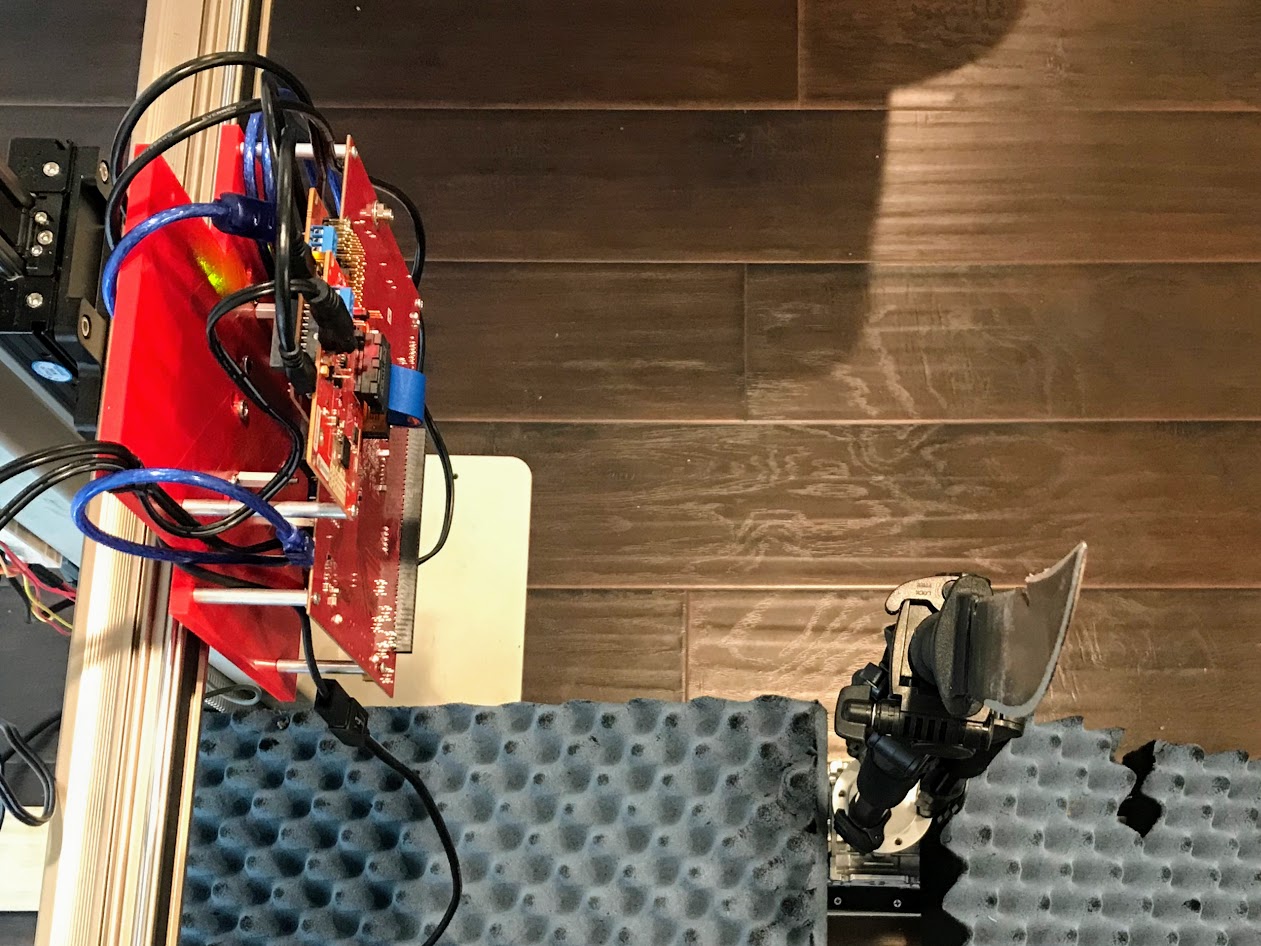}
		\caption{Knife Parallel}
		\label{fig:scanner_parallel}
	\end{subfigure}%
	\begin{subfigure}{.5\linewidth}
		\centering
		\includegraphics[width=1\linewidth]{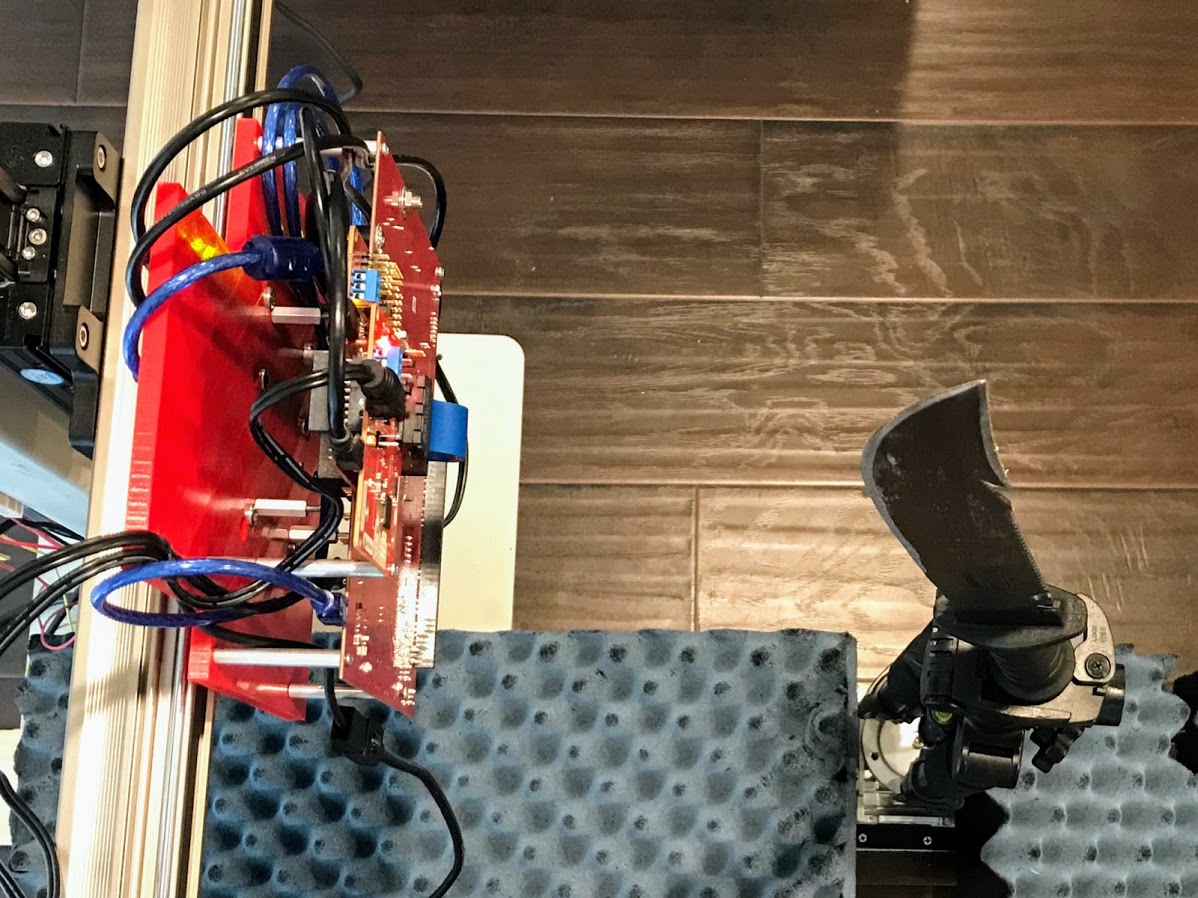}
		\caption{Knife Perpendicular}
		\label{fig:scanner_perp}
	\end{subfigure}
	\caption{Knife parallel (a) and perpendicular (b) to scanning plane for rectilinear SAR scans.}
	\label{fig:MIMO_SAR_System_Configuration}
\end{figure}

The R-ISAR regime is more suitable for near-field target scanning than rectilinear (planar) SAR regime because the target is scanned by the radar from all sides. To demonstrate the advantage of R-ISAR over SAR, a rectilinear 2-D scan is performed across the x and y axes of the mechanical scanner, without rotating the knife, using the parameters in Table \ref{table_SARradar_parameters}, where $D^x_S$ is the size of the horizontal scan. Two scenarios are proposed. First, the knife is scanned with its blade parallel to the scanning plane; then, it is scanned again with the blade perpendicular to the scanning plane. Both configurations are shown in Fig. \ref{fig:MIMO_SAR_System_Configuration}. When the knife is parallel, the resulting image is a high quality reconstruction resembling the blade (Fig. \ref{fig:knife_MIMO_SAR_Parallel_mip}). But when it is perpendicular, the image of the blade is quite poor (Fig. \ref{fig:knife_MIMO_SAR_Perp_mip}).

\begin{table} [h]
	\caption{\label{table_SARradar_parameters}MIMO SAR Radar Parameters}
	\centering
	\begin{tabular}{c c c c c}
		\hline
		$D^S_x$ & $\Delta_x$ & $N_y$ & $D^S_y$ & $\Delta_y$  \\ [0.5ex] 
		\hline\hline
		450 mm & 0.5 mm & 64 & 484.8 mm & $2\lambda$ \\ 
		\hline
	\end{tabular}
\end{table}

Examining all four images in Fig. \ref{fig:knife_MIMO} qualitatively, the image quality of the MIMO R-ISAR image (Fig. \ref{fig:knife_MIMO_mip}) appears to be the same, if not better, than that of the SISO R-ISAR scan (Fig. \ref{fig:knife_SISO_mip}). For the MIMO-SAR (planar) scans, while the image quality of the knife for the perpendicular scan (\ref{fig:knife_MIMO_SAR_Parallel_mip}) is comparable to the MIMO R-ISAR image, targets are not always ideally parallel to the scanning plane. When the knife is scanned perpendicular to the scanning plane (Fig. \ref{fig:knife_MIMO_SAR_Perp_mip}), the quality of the reconstructed image of the blade is degraded substantially since the cross range resolution of the SAR regime is significantly less than its R-ISAR counterpart. For the R-ISAR regime, the image quality is independent of the orientation of the object with respect to the rotation axis since the target is rotated a full 360$^\circ$. Therefore, the R-ISAR regime captures the scanned target both parallel and perpendicular at different times throughout the rotation. Lastly, considering time required for each scanning regime, as shown in table \ref{table_scanning_times}, the R-ISAR mode offers the fastest scanning and best image performance. By using the algorithm proposed in this paper, we were able to drastically reduce the scanning time while maintaining the computational efficiency of the monostatic algorithms.

\begin{table} [h]
	\caption{\label{table_scanning_times}Scanning Times}
	\centering
	\begin{tabular}{c c c}
		\hline
		SISO R-ISAR & MIMO R-ISAR & MIMO SAR  \\ [0.5ex] 
		\hline
	    1032 s & 8202 s & 1740 s\\
		\hline
	\end{tabular}
\end{table}
		
\section{Conclusion}
\label{Sec_conclusion}
In this paper, we developed an efficient MIMO rotational-ISAR 3-D holographic imaging system based on the single pixel polar formatting algorithm and multistatic-to-monostatic conversion. The algorithm successfully pairs the scanning efficiency of MIMO systems with the computational efficiency of monostatic reconstruction algorithms. Additionally, we developed a complete, robust 3-D imaging system consisting of a vertically scanned MIMO radar and a rotator to rotate the target object. Our system fully integrates scanning scenario setup, data collection and calibration, algorithm implementation, and image inspection for a complete, efficient 3-D holographic imaging platform. Using this prototype system, high-resolution images are captured to demonstrate the effectiveness of this system for 3-D scene reconstruction and verify the MIMO R-ISAR algorithm's performance in comparison to its SISO R-ISAR and MIMO-SAR counterparts. Additionally, the MIMO R-ISAR regime is shown to outperform the MIMO-SAR regime with improved range and cross-range resolution, deeming R-ISAR more suitable for near-field imaging applications. The algorithm and system demonstrate high-performance near-field imaging using MIMO rotational inverse synthetic aperture radar.

\section*{Acknowledgment}
This work is supported by Semiconductor Research Corporation (SRC) task 2712.029 through The University of Texas at Dallas' Texas Analog Center of Excellence (TxACE).

\bibliography{MIMO_ISAR_RadarConf20_Paper}
\bibliographystyle{IEEEtran}

\end{document}